\documentstyle[12pt,aaspp4]{article}
\begin{document}

\newcommand{\up}[1]{\ifmmode^{\rm #1}\else$^{\rm #1}$\fi}
\newcommand{\zdot}{\makebox[0pt][l]{.}}
\newcommand{\upd}{\up{d}}
\newcommand{\uph}{\up{h}}
\newcommand{\upm}{\up{m}}
\newcommand{\ups}{\up{s}}
\newcommand{\arcd}{\ifmmode^{\circ}\else$^{\circ}$\fi}
\newcommand{\arcm}{\ifmmode{'}\else$'$\fi}
\newcommand{\arcs}{\ifmmode{''}\else$''$\fi}

\title{The Araucaria Project. The Distance to the Sculptor Group
Galaxy NGC 55 from a Newly Discovered Abundant Cepheid Population.
\footnote{Based on  observations obtained with the 1.3~m Warsaw
telescope at Las Campanas Observatory, Chile.
}
}
\author{Grzegorz Pietrzy{\'n}ski}
\affil{Universidad de Concepci{\'o}n, Departamento de Fisica, Astronomy
Group,
Casilla 160-C,
Concepci{\'o}n, Chile}
\affil{Warsaw University Observatory, Al. Ujazdowskie 4,00-478, Warsaw,
Poland}
\authoremail{pietrzyn@hubble.cfm.udec.cl}
\author{Wolfgang Gieren}
\affil{Universidad de Concepci{\'o}n, Departamento de Fisica, Astronomy Group, 
Casilla 160-C, 
Concepci{\'o}n, Chile}
\authoremail{wgieren@astro-udec.cl}
\author{Igor Soszy{\'n}ski}
\affil{Universidad de Concepci{\'o}n, Departamento de Fisica, Astronomy
Group, Casilla 160-C, Concepci{\'o}n, Chile}
\affil{Warsaw University Observatory, Aleje Ujazdowskie 4,
PL-00-478,Warsaw, Poland}
\author{Andrzej Udalski}
\affil{Warsaw University Observatory, Aleje Ujazdowskie 4, PL-00-478,
Warsaw,Poland}
\authoremail{udalski@astrouw.edu.pl}
\author{Fabio Bresolin}
\affil{Institute for Astronomy, University of Hawaii at Manoa, 2680 Woodlawn 
Drive, 
Honolulu HI 96822, USA}
\authoremail{bresolin@ifa.hawaii.edu}
\author{Rolf-Peter Kudritzki}
\affil{Institute for Astronomy, University of Hawaii at Manoa, 2680 Woodlawn 
Drive, Honolulu HI 96822, USA}
\authoremail{kud@ifa.hawaii.edu}
\authoremail{agarcia@astro-udec.cl}
\author{Ronald Mennickent}
\affil{Universidad de Concepci{\'o}n, Departamento de Fisica, Astronomy
Group, Casilla 160-C,
Concepci{\'o}n, Chile}
\authoremail{rmennick@astro-udec.cl}

\author{Alistair Walker}
\affil{Cerro Tololo Inter-American Observatory, Casilla 603, La Serena, Chile}
\authoremail{awalker@ctio.noao.edu}
\author{Alejandro Garcia}
\affil{Universidad de Concepci{\'o}n, Departamento de Fisica, Astronomy
Group, Casilla 160-C, Concepci{\'o}n, Chile}
\authoremail{agarcia@astro-udec.cl}
\author{Olaf  Szewczyk}
\affil{Warsaw University Observatory, Aleje Ujazdowskie 4, PL-00-478,
Warsaw,Poland}
\authoremail{szewczyk@astrouw.edu.pl}
\author{Micha{\l} Szyma{\'n}ski}
\affil{Warsaw University Observatory, Aleje Ujazdowskie 4, PL-00-478,
Warsaw, Poland}
\authoremail{msz@astrouw.edu.pl}
\author{Marcin Kubiak}
\affil{Warsaw University Observatory, Aleje Ujazdowskie 4, PL-00-478,
Warsaw, Poland}
\authoremail{mk@astrouw.edu.pl}
\author{{\L}ukasz Wyrzykowski}
\affil{Warsaw University Observatory, Aleje Ujazdowskie 4, PL-00-478,
Warsaw,Poland}
\authoremail{wyrzykow@astrouw.edu.pl}

\begin{abstract}
We have detected, for the first time, Cepheid variables in the Sculptor Group SB(s)m
galaxy NGC 55. From wide-field images obtained in the optical V and I bands
during 77 nights in 2002-2003, we have found 143 Cepheids with periods
ranging from 5.6 to 175.9 days. 133 of these objects have periods longer than
10 days, making NGC 55 to-date the galaxy with the largest known number of long-period
Cepheids in the Sculptor Group. We construct period-luminosity
relations from our data and obtain distance moduli corrected for the small foreground reddening
to NGC 55 of 26.79 $\pm$ 0.04 mag (internal error) in V, 26.66 $\pm$ 0.03 mag in I
and 26.40 $\pm$ 0.05 mag in the
reddening-independent V-I Wesenheit index. The trend of increasing distance moduli with
shorter wavelength hints at the existence of significant reddening intrinsic to NGC 55
which affects the measured Cepheid magnitudes. From our data, we determine the intrinsic
mean reddening of the Cepheids in NGC 55 as E(B-V) = 0.102 mag which
brings the distance determinations
from the different bands into excellent agreement. Our best distance estimate
for NGC 55 from the present optical Cepheid photometry is 26.40 mag $\pm$ 0.05 mag (internal error)
$\pm$ 0.09 mag (systematic error). This value is tied to an assumed LMC distance of 18.50 mag.
Our quoted systematic error of the present NGC 55 Cepheid distance does not take into account the
current uncertainty on the distance of the fiducial LMC galaxy itself.

Within the small respective uncertainties, the Sculptor Group galaxies NGC 55 and NGC 300
are at the same distance of 1.9 Mpc, strengthening the case for a physical association
of these galaxies.
\end{abstract}


\section{Introduction}
In our ongoing Araucaria Project, we are improving the accuracy of several of the
most important stellar distance indicators with the aim to measure the distances to nearby
galaxies, out to about 10 Mpc, with an accuracy of 3 percent with respect to the Large
Magellanic Cloud (LMC). The feasibility to reach this accuracy using 
a combination of optical and near-infrared photometry of {\it Cepheid variables}
has already been demonstrated for
the Local Group irregular galaxies IC 1613 (Pietrzy{\'n}ski et al. 2006a), NGC 6822 (Gieren et al. 2006;
Pietrzy{\'n}ski et al. 2004)
and NGC 3109 (Pietrzy{\'n}ski et al. 2006b, Soszy{\'n}ski et al. 2006). Our list of targets
also includes four of the five principal late-type galaxies in the loose neighboring group
of galaxies in Sculptor. For NGC 300, a near face-on spiral member of the Sculptor Group 
at a distance of 1.88 Mpc,
we were able to achieve the same 3 percent accuracy for its distance determination from the
Cepheid approach (Gieren et al. 2005a). For all these galaxies so far studied in the Araucaria
Project, we found clear evidence for the existence of substantial intrinsic dust absorption, in addition
to the foreground absorption produced inside the Milky Way. This result underlines the importance
to use infrared data to obtain the most reliable distance estimates, eliminating intrinsic
reddening as the dominating source of systematic error in optical photometric studies. Our previous
studies also show that in the case of Cepheids, the reddening-independent {\it Wesenheit index}
is able to provide a distance result which is much less affected by reddening than the
distance estimates in the V and I bands, provided excellent optical light curves of the
variables are available. A Cepheid distance derived from the (V-I) Wesenheit index (for
a definition see our previous papers cited above) is therefore a good approximation
to the more accurate distance which can be obtained from near-infrared data, and allows an important
consistency check on the distance result for a given galaxy from infrared photometry of its Cepheid
variables.

NGC 55, our target galaxy studied in this paper, is a highly inclined late-type galaxy classified as
SB(s)m in the NASA Extragalactic Database. Although its high inclination makes it difficult
to study its morphology, NGC 55 resembles to some extent the LMC but is more massive. From 
color images as the one shown in Gieren et al. (2005b), it is evident that NGC 55 contains an abundant
population of young blue stars and should therefore contain a sizable population of Cepheids, which have
so far not been searched for in this galaxy. The few previous distance determinations which have been
attempted for NGC 55 have yielded an unsatisfactorily large range, from 1.34 Mpc from carbon stars
(Pritchet et al. 1987) to 2.3 Mpc from planetary nebulae (Van de Steene et al. 2006). In between
falls the determination of 1.8 Mpc obtained by Karachentsev et al. (2003)
who used the Tully-Fisher
relation, tied to distances of other dwarf galaxies in the Sculptor Group measured with 
the tip of the red giant branch method. All these distance determinations are likely to be plagued by rather
large systematic uncertainties. For instance, the recent determination of Van de Steene et al. (2006)
from the PNLF method depends rather sensitively on the assumed oxygen abundance of the planetary
nebulae in NGC 55 used in the construction of the luminosity function, which is quite uncertain at the
present time (e.g. Tuellmann et al. 2003). The detection of a sizable population of Cepheid variables
in NGC 55 has therefore a strong potential to improve the situation and provide a more accurate
distance determination for this galaxy which we need in the context of the principal science goal
of the Araucaria Project, viz. to determine the effect of metallicity on the physical parameters
used as a distance indicator for the different stellar standard candles we study. We remark here that
we are obtaining, in a parallel effort, chemical abundances for oxygen and some other elements
from the spectra of a large number of blue supergiant stars in NGC 55 which we have obtained with the
ESO VLT. These abundances will not only provide a mean metallicity for NGC 55 but also allow to determine
the abundance {\it gradient} in its disk if existent, as we were able to do in the case of NGC 300 (Urbaneja
et al. 2005).

The purpose of this paper is to present the results of an extensive wide-field imaging survey
for Cepheid variables in NGC 55 which we conducted in 2002 and 2003 at the Warsaw
1.3-m telescope at Las Campanas Observatory. As we will show in the following sections, we have
discovered a surprisingly large population of long-period Cepheids in this galaxy. We will describe
our observations, data reductions and calibrations in section 2, and will present the catalog of photometric
properties of the Cepheid variables in NGC 55 in section 3. In section 4 we will determine the period-luminosity
relations in the V, I and Wesenheit bands from our data and provide a new distance determination
to NGC 55. We will discuss our results in section 5, and summarize our conclusions in section 6.

\section{Observations,  Reductions and Calibrations}
All the  data presented in this paper were collected with the Warsaw 1.3-m 
telescope at Las Campanas Observatory. The telescope was equipped with 
a mosaic 8k $\times$ 8k detector, with a field of view of about 35 $\times$ 35 
arcmin and a scale of about 0.25 arcsec/pix. For more  instrumental
details on this camera, the reader is referred to the OGLE  website:
{\it http://ogle.astrouw.edu.pl/\~{}ogle/index.html}.
V and I band  images of NGC 55 were secured during 77 different nights.
The exposure time was set to 900 seconds in both filters.
Most of the observations were obtained in 2002 between July 3 and November 21.
Additional observations were made during three photometric nights in May 2003 
in order to accurately calibrate the data and improve the periods of the
detected variable stars. 

Preliminary reductions (i.e. debiasing and flatfielding)  were 
done with the IRAF\footnote{IRAF is distributed by the
National Optical Astronomy Observatories, which are operated by the
Association of Universities for Research in Astronomy, Inc., under cooperative
agreement with the NSF.} package. Then, the PSF photometry was obtained 
for all stars in the same manner as in Pietrzy{\'n}ski, Gieren and
Udalski (2002). Independently, the data were reduced with the OGLE III pipeline 
based on the image subtraction technique (Udalski 2003; Wo{\'z}niak 2000).  

In order to calibrate our photometry onto the standard system 
our target was monitored during three photometric nights
(2003 May 11, 17 and 28) together with 
some 25 standard stars from the Landolt fields spanning a wide range of
colors ( -0.14 $<$ V-I $<$ 1.43), and observed at very different
airmasses. Since in principle the transformation equations for each of the eight
chips may have different color coefficients and zero points, the selected
sample of standard stars was observed on each of the individuial chips, and
transformation coefficients  were derived independently for each chip, 
on each night. Comparison of the photometry obtained on the different nights
revealed that the internal accuracy of zero points in both V and
I bands is better than 0.02 mag. 

To correct the variation of the zero points in V and I
over the mosaic, the "correction maps" established by Pietrzy{\'n}ski 
et al. (2004) were used. These maps were already applied to correct 
photometry obtained in the field of NGC 6822 (Pietrzy{\'n}ski et al. 
2004) and NGC 3109 (Pietrzy{\'n}ski et al. 2006b) with the same camera. Comparison with 
other studies revealed that these maps allow to correct for
the zero point variations across the mosaic field in a way that the
residuals do not exceed 0.02-0.03 mag. 

Two fields in the SMC which contain a huge number of uniformly
distributed  stars 
with very accurate photometry established by the OGLE team (Udalski 
et al. 1999) were monitored on all three nights together with our target and 
standard fields, and were calibrated in the same manner as the NGC 55 
observations. Comparison of OGLE II and our photometry 
shows that on average the difference in the zero points between 
these two photometric sets is smaller than 0.02 mag on each chip. 

Alcaino and Liller (1983) established a sequence of secondary standards in the
field of NGC 55. In our database 7 stars were found to be common to this
sequence (e.g. the remaining ones were either saturated or located outside 
our field of view). The comparison of our V band photometry for these objects with
that of Alcaino and Liller, which is presented in Table 1, demonstrates that the zero points 
of the two datasets agree to within 0.02 mag.  

Summarizing the previous discussion, we may conclude that all internal and external checks 
we made indicate that the zero points of our data are accurate to 0.02 mag 
in both V and I, and that their possible variations over the mosaic 
fields are smaller than 0.02 mag.

\section{The Catalog of Cepheids}
All stars with photometry in NGC 55 were searched for photometric variations with
periods between 0.2 and 200 days, using the analysis of variance algorithm 
(Schwarzenberg-Czerny 1989). In order to distinguish Cepheids from other types of
variable stars, we used the same criteria given in our initial paper reporting on the
discovery of Cepheids in NGC 300 (Pietrzy{\'n}ski et al. 2002). The light curves
of all Cepheid candidates were approximated by Fourier series of order 4. We then
rejected those objects with V amplitudes smaller than 0.4 mag, the approximate 
lower limit amplitude for classical Cepheids. This procedure also helps in screening
our Cepheid sample from the inclusion of strongly blended objects. For the variable stars
passing our selection criteria, mean V and I magnitudes were derived by integrating
their light curves which had been previously converted onto an intensity scale, and converting
the results back to the magnitude scale. Because of the very good quality and phase coverage of 
the light curves, the statistical accuracy of the derived mean magnitudes was typically
0.01 mag for the brighter variables, increasing to about 0.03 mag for the faintest Cepheids in
our catalog. We ended up with a list of 143 bona fide classical Cepheids with periods between
175.9 and 5.6 days. The number of long-period Cepheids in NGC 55 is surprisingly large, with
133 of the detected Cepheids having periods longer than 10 days. This makes NGC 55 even richer
in its long-period Cepheid population than NGC 300, the other Sculptor galaxy we had previously
surveyed for Cepheids (Gieren et al. 2004).  There is certainly a note of caution as to
the true Cepheid nature of the 3 objects with periods in excess of 100 days; the time baseline
of our photometry is possibly too short to establish their Cepheid character beyond any
doubt. Future observations will clarify this question. For this reason, we will exclude
these variables in our distance determination in the next section, as we did in the case
of the Local Group galaxy NGC 6822 in which we had also found a number of very long-period
Cepheid-like variables (Gieren et al. 2006; Mennickent et al. 2006).

Table 2 gives the individual photometric V and I observations for all Cepheid variables
we found in NGC 55. The full Table 2 is available in electronic form. In Table 3, we present their
identifications, equatorial coordinates, periods, times of maximum brightness in V, and their
mean intensity magnitudes in the V, I and W bands. The accuracy of the periods is typically
about $10^{-3}*P$ days, good enough to be no significant source of error in the
construction of the period-luminosity relations discussed in the next section.

In Fig. 1, we show a mosaic image of NGC 55 from our data with the positions of the detected
Cepheids overplotted. Fig. 2 presents light curves for some of the Cepheids which are
representative in quality for other variables in our catalog of similar periods, showing
that the mean magnitudes of the Cepheids can indeed be derived quite accurately from our data even for
the shortest-period and faintest Cepheids in our sample. In Fig. 3 we show the locations
of the Cepheids in the V, V-I color-magnitude diagram for NGC 55 which we
have constructed from our data. The diagram shows that the Cepheids indeed do fill the expected
zone in the CMD delineating the Cepheid instability strip, presenting supporting evidence
that our classification of the objects in Table 3 as Cepheids seems correct in all cases.

\section{Period-Luminosity Relations and Distance Determination}
Although a more accurate distance to NGC 55 will be derived once near-infrared
follow-up photometry will become available for a substantial subsample of the Cepheids in
our catalog, we can already clearly improve on the existing distance determinations
for NGC 55 with the optical Cepheid photometric data we have provided in this paper. In Figs. 4-6,
we show the period-luminosity (PL) diagrams obtained from the data in our Cepheid
catalog for the 130 stars with periods in the range from 10-100 days. PL relations are plotted
for the V, I and W magnitudes, where the (V-I) Wesenheit magnitude is
defined in its usual way (see previous papers in the Araucaria Project). The PL
relation in V is well defined although with somewhat larger scatter than
in the case of NGC 300 at a similar distance (Pietrzy{\'n}ski et al. 2002; Gieren et al. 2004).
This is likely due to an increased size of differential reddening in NGC 55 due
to its high inclination with respect to the line of sight, as compared to NGC 300
which is oriented almost face-on. As expected if differential reddening indeed causes much 
of the scatter in Fig. 4,
the I-band PL relation in Fig. 5 has a clearly smaller dispersion. While it should be
expected that the W-band PL relation has an even smaller dispersion due to its expected
insensitivity to reddening, Fig. 6 shows that this is not the case and that the
dispersion is instead rather similar to the one in the V-band PL relation. This is probably
because the photometric accuracy of the data especially for the shorter-period Cepheids
is not quite as high as needed for the W-band technique to show its full power.
It is also possible that crowding and blending of the Cepheids with bright unresolved companion stars
 in this high-inclination galaxy is a more
significant problem than in the other galaxies we have studied so far, invalidating
the power of the Wesenheit approach to some extent. This has to be kept in mind
when discussing the accuracy of our present distance determination.
 
In order to obtain the distance to NGC 55, we have carried out
least-squares fits to a line to the various PL diagrams. These free fits yield the following 
slopes for the PL relations: -2.31 $\pm$ 0.18, -2.56 $\pm$ 0.12 
and -2.85 $\pm$ 0.20 in V, I and ${\rm W_{\rm I}}$, respectively. These slope
values are somewhat shallower than the corresponding slopes for the LMC Cepheid sample studied
by the OGLE Project, but they are still consistent with the OGLE LMC PL relation slopes
of -2.775 in V, -2.977 in I and -3.300 in ${\rm W_{\rm I}}$ (Udalski 2000) at the level of 2 $\sigma$.  
We therefore keep our standard procedure adopted in our previous papers in this series
to use the OGLE LMC slopes to derive the distance of our target galaxies relative to the LMC.
Forcing the slopes to the respective OGLE values leads to the following equations
for the NGC 55 Cepheid PL relations:\\

V = -2.775 log P + (25.401 $\pm$ 0.038) mag\\

I = -2.977 log P + (24.778 $\pm$ 0.026) mag\\

${\rm W}_{\rm I}$ = -3.300 log P + (23.765  $\pm$ 0.045) mag \\

Adopting 18.50 mag for the LMC distance modulus as we did in our previous papers, and using
E(B-V) = 0.014 mag as the foreground 
reddening towards NGC 55 (Schlegel et al. 1998) together with the reddening law of the same authors
( ${\rm A}_{\rm V}$ = 3.24 E(B-V), ${\rm A}_{\rm I}$ = 1.96
E(B-V)), we obtain the following distance moduli for NGC 55 from the
three different bands:\\

$(m-M)_{0}$ (${\rm W}_{\rm I}$) = 26.40 $\pm$ 0.05 mag \\    

$(m-M)_{0}$ (I) = 26.66 $\pm$ 0.03 mag\\

$(m-M)_{0}$ (V) = 26.79 $\pm$ 0.04 mag\\

These values show the same tendency of increasing distance moduli towards the bands
with the higher reddening sensitivity that we have seen for the other galaxies
previously studied in this series,
suggesting the existence of significant internal reddening in NGC 55 in addition
to the foreground reddening produced in the Milky Way.

This is of course no surprise because high-resolution images of NGC 55 clearly reveal
the existence of significant dust absorption
in the regions where the young, blue stars are concentrated.
Since the difference of the distance moduli in V and I above is identical to the reddening
E(V-I), we obtain from our data E(V-I) = 26.79 - 26.66 = 0.13 mag. With E(B-V) = E(V-I)/1.28,
this translates in a (B-V) color excess of E(B-V) = 0.102 mag, which then leads to an
extinction-corrected true distance modulus of $(m-M)_{0}$ = 26.46 mag. This value agrees
within the combined 1 $\sigma$ uncertainty with the true distance modulus derived from
the Wesenheit index. Unless follow-up near-infrared data for the Cepheids in NGC 55 become
available which will allow a more accurate determination of the mean reddening of the
Cepheids due to the longer wavelength baseline, we will adopt, as our current best value
from optical VI photometry, 26.40 mag as the true distance modulus of NGC 55.

\section{Discussion}
Our adopted distance to NGC 55 from optical photometry in the V and I bands of 130
long-period Cepheid variables discovered in the present survey is 26.40 mag, corresponding to
1.91 Mpc. This value is practically identical to the distance of NGC 300 of (1.88 $\pm$ 0.06) Mpc
(Gieren et al. 2005a) , strongly supporting the idea that NGC 300 and NGC 55 form a gravitationally
bound system as previously suggested by Whiting (1999), and others. The intrinsic uncertainty
of the distance modulus of NGC 55 derived from the (V-I) Wesenheit index is $\pm$ 0.05 mag,
and is basically determined by the accuracy of our photometry. There are additional sources of
systematic error which may affect this result. One of them is the effect of blending of some Cepheids
in our sample
with relatively bright stars, too close to the Cepheids to be resolved in our images.
In a previous study of NGC 300 in which we could compare ground-based photometry of a large
number of Cepheids to BVI photometry of the same variables obtained with the {\it Hubble
Space Telescope} (Bresolin et al. 2005), we were able to show that the effect blending with
unresolved bright stars has on the distance modulus derived from the ground-based images is less than 0.04 mag.
We think that the effect of blending on our present distance result for NGC 55 should be of the same order
given that the galaxy has the same distance as NGC 300, and the resolution of our NGC 55 images
is very similar to the one of the images we used in our distance work for NGC 300.
The higher inclination of NGC 55
with respect to the line of sight however might render the effect due to blending somewhat more severe,
as possibly indicated by the slightly enhanced dispersion of the V-band PL relation for NGC 55 (Fig. 4)
as compared to NGC 300 (Gieren et al. 2004). We estimate that blending with unresolved bright companion
stars might introduce a systematic uncertainty of $\pm$ 0.07 mag, in the case of NGC 55.

Another possible source of systematic uncertainty on our distance result is the slopes of the fiducial
LMC Cepheid PL relations which we have adopted from the extensive work of the OGLE Project (Udalski 2000).
Recent work by Ngeow and Kanbur (2006; see also references given therein) suggest that the LMC
Cepheid PL relations in V and I are likely to show significant breaks at a period close to 10 days, leading
to slightly different slopes of the PL relations valid for periods shorter, and longer than 10 days.
The effect is very subtle and according to Ngeow and Kanbur (2006) affects distance moduli determined
from the V and I PL relations by less than 0.03 mag.
We will assume here that the systematic uncertainty
of our NGC 55 distance modulus due to the uncertain slope of the fiducial PL relation in the LMC
is $\pm$ 0.03 mag.

Another effect which might systematically affect our distance result is the possible metallicity
dependence of the Cepheid PL relation. If such an effect is significant, the use of a PL relation calibrated
for LMC metallicity for a Cepheid population whose metallicity is quite different from the LMC Cepheids
would systematically affect the distance result.
A number of recent theoretical and empirical studies have tried
to trace down the effect of Cepheid metal abundances on their absolute magnitudes in different wavelength
ranges (e.g. Storm et al. 2004; Saha et al. 2006), but unfortunately so far no clear-cut and accurate
results have been obtained.
A thorough discussion of the metallicity effect on the Cepheid PL relation
 has very recently been given by Mottini et al. (2006).
Their own results obtained from relating the PL relation residuals for individual Cepheids to their
metallicities measured from high-resolution spectra
are consistent with a null effect of Cepheid iron abundance on the PL relation zero points
in both the optical V and the near-infrared K band. It is clear, however, that further
work on this very important question is needed which will eventually provide more accurate results than all
previous work published over the past years.
This is one of the principal goals of our
Araucaria Project. What can be said at the present time is that from the results published
over the past 8 years (see Mottini et al. 2006)
it appears that the metallicity effect is small, not exceeding a few percent in any of the standard optical
and near-IR photometric bands. This seems true both for the PL relation zero
points (Pietrzynski and Gieren 2006), and for
their slopes in different bands (e.g. Gieren et al. 2005c). We adopt $\pm$ 0.04 mag as an additional
systematic uncertainty on our present distance modulus for NGC 55 from the metallicity effect.

Probably the largest systematic uncertainty on our present distance result for NGC 55 is the size
of the reddening intrinsic NGC 55 which affects our Cepheid sample. The use of the Wesenheit
index in this paper, and the consistency of the distance moduli from V, I and W using the reddening value
derived from our data makes us expect that
the effect of the uncertain intrinsic reddening on our adopted distance modulus is not larger than
a few percent, however. It is clear that our planned follow-up distance determination for NGC 55
from accurate near-infrared photometry of a significant subsample of the Cepheids in our catalog
will eliminate reddening as a significant source of error, as we could demonstrate in our
previous Cepheid near-infrared studies in the Araucaria Project, and allow us to reduce the
total systematic uncertainty on the distance of NGC 55 to a figure close to 3\%. Moving our
distance work to the near-infrared will also reduce the effect of blending because
most stars which are bright enough to contaminate the fluxes of Cepheid variables at a significant
level, including the secondary components in Cepheid binary systems,
 are expected to be blue stars which affect the Cepheid fluxes in optical bands more strongly
than in the near-infrared (e.g. Gieren 1982; Szabados 1996).

Summarizing this discussion, we estimate that the total systematic uncertainty on our current
distance modulus for NGC 55 from the various factors we have discussed is in the order of $\pm$ 0.09 mag.
Our adopted true distance modulus to NGC 55 is therefore 26.40 mag $\pm$ 0.05 mag (internal error)
$\pm$ 0.09 mag (systematic error). We note that this result is tied to an {\it assumed} distance
modulus of 18.50 of the LMC and does not include the current uncertainty on the LMC distance, which is
probably in the order of $\pm$ 0.10 mag as well. Since all the distances we are obtaining in the
Araucaria Project are tied to the same assumed LMC distance of 18.50 mag, however, the {\it relative} distances
between our target galaxies are not affected.

\section{Conclusions}
We have carried out an extensive search for Cepheid variables in NGC 55 which has resulted in the
discovery of 143 bona fide Cepheids. We provide the periods of these objects and their mean magnitudes
in the optical V and I bands, as well as their accurate positions. From 130 of these variables
with periods between 10 and 100 days we obtain the period-luminosity relations in V and I, and for the
reddening-independent (V-I) Wesenheit index. The distance moduli derived from these data
demonstrate that apart from the very small foreground reddening, there is significant additional reddening
which is produced inside NGC 55, which we determine as E(B-V) = 0.102 mag from our data.
Our best adopted value
for the distance modulus of NGC 55 from optical photometry of its large population of Cepheid
variables is 26.40 mag $\pm$ 0.05 mag (intrinsic) $\pm$ 0.09 mag (systematic). The systematic
uncertainty includes the estimated contributions from the effect of metallicity on the PL relation,
the current uncertainty on the slope of the fiducial LMC Cepheid PL relation we have used
in our distance determination, the effect
of blending of the Cepheids in our sample with bright companion stars which are not resolved
in our images, accuracy of the photometric zero points,  and of dust absorption internal to NGC 55
affecting the optical Cepheid magnitudes.
As in the previous papers of the Araucaria Project, our adopted distance modulus of our
present target galaxy is tied to an assumed LMC true distance modulus of 18.50. Our estimated
0.09 mag systematic uncertainty of the distance modulus of NGC 55 does not contain the
additional systematic uncertainty of the adopted value of the LMC distance.

The Cepheid distance derived in this paper for NGC 55 is more accurate
than the other distance determinations for this nearby galaxy available in the literature which
we discussed in the Introduction of this paper.
In the near future, we expect to improve the accuracy of the distance determination to NGC 55 even further
through near-IR photometry of a subsample of the Cepheids in our catalog,
which will reduce the current uncertainties in our optical study due to
reddening and blending.

Within the errors, the distance to NGC 55 from its Cepheid variables is {\it identical} to the Cepheid distance
of 1.88 Mpc to NGC 300 (Gieren et al. 2005a) which is located close to NGC 55 in the sky.
This strengthens the case for a physical association of NGC 55 and NGC 300, as previously suggested
by a number of authors.

\acknowledgments
We are grateful to Las Campanas Observatory for the support provided
to our observers, 
and to the CNTAC for providing the large amounts of 
telescope time which were necessary to complete this project.
We also would
like to thank the anonymous referee for interesting comments and suggestions.
GP, WG, RM and AG  gratefully acknowledge 
financial support for this
work from the Chilean Center for Astrophysics FONDAP 15010003. 
WG also acknowledges support
from the Centrum fuer Internationale Migration und Entwicklung in
Frankfurt/Germany
who donated the workstation on which a substantial part of the data 
reduction and analysis for this project was carried out.   
Support from the Polish KBN grant No 2P03D02123 and BST grant for 
Warsaw University Observatory is also acknowledged.

\begin{deluxetable}{c c c c c c c c}
\tablecaption{Comparison of our V band photometry}
\tablehead{
\colhead{ID} & \colhead{V [mag]} & V [mag] & Remarks \\
\colhead{}& \colhead{Alcaino and Liller}  & this paper & \\
}
\startdata
V &  16.05 &   16.054& \\
W &  16.20 &   16.212& \\
X &  16.95 &   17.033& Variable star\\
T &  15.79 &   15.796& \\
K &  14.47 &   14.448& \\
N &  14.90 &   14.919& \\
O &  14.91 &   14.916& \\
\enddata
\end{deluxetable}

\begin{deluxetable}{c c c c c c c c}
\tablecaption{Cepheids in NGC 55}
%\tablewidth{0pt}
\tablehead{
\colhead{ID} & \colhead{RA} & \colhead{DEC}  & \colhead{P} & \colhead{ ${\rm
T}_{0}$} &
\colhead{$<V>$} & \colhead{$<I>$} & \colhead{$<W_{\rm I}>$}\\
 & \colhead{(J2000)} & \colhead{(J2000)}  &
\colhead{ [days]} &  &
\colhead{[mag]} & \colhead{[mag]} & \colhead{[mag]}
}
\startdata
cep001 & 0:14:12.95 & -39:08:41.7 & 175.9086 & 2452808.50787 &  19.249 &  18.409 &  17.107 \\ 
cep002 & 0:15:12.00 & -39:12:18.0 & 152.0943 & 2452746.27372 &  19.562 &  18.613 &  17.142 \\ 
cep003 & 0:14:36.57 & -39:11:08.9 & 112.6964 & 2452800.19911 &  20.177 &  19.131 &  17.510 \\ 
cep004 & 0:15:14.28 & -39:13:17.4 &  97.7291 & 2452823.21391 &  20.536 &  19.288 &  17.354 \\ 
cep005 & 0:15:10.11 & -39:12:26.0 &  85.0550 & 2452769.04780 &  20.835 &  19.456 &  17.319 \\ 
cep006 & 0:15:47.32 & -39:16:10.7 &  76.6842 & 2452797.49527 &  20.101 &  19.147 &  17.668 \\ 
cep007 & 0:15:40.34 & -39:15:08.0 &  75.8328 & 2452788.65035 &  20.336 &  19.510 &  18.230 \\ 
cep008 & 0:14:45.35 & -39:13:17.4 &  74.6393 & 2452828.22484 &  20.124 &  19.253 &  17.903 \\ 
cep009 & 0:15:13.99 & -39:12:33.4 &  73.5323 & 2452782.28141 &  20.912 &  19.706 &  17.837 \\ 
cep010 & 0:14:06.59 & -39:08:11.6 &  66.8528 & 2452788.45946 &  20.356 &  19.464 &  18.081 \\ 
cep011 & 0:13:58.27 & -39:08:22.8 &  62.7910 & 2452799.95331 &  20.432 &  19.568 &  18.229 \\ 
cep012 & 0:15:14.93 & -39:12:52.5 &  62.3186 & 2452811.59034 &  20.834 &  19.665 &  17.853 \\ 
cep013 & 0:14:12.72 & -39:09:27.2 &  57.1544 & 2452802.83374 &  20.425 &  19.297 &  17.549 \\ 
cep014 & 0:15:56.86 & -39:15:23.6 &  57.0377 & 2452840.41529 &  20.228 &  19.252 &  17.739 \\ 
cep015 & 0:14:54.55 & -39:12:49.4 &  56.9074 & 2452832.44085 &  20.418 &  19.550 &  18.205 \\ 
cep016 & 0:14:37.74 & -39:10:55.9 &  53.1177 & 2452834.23720 &  20.746 &  19.595 &  17.811 \\ 
cep017 & 0:15:00.36 & -39:12:25.9 &  48.7260 & 2452809.41765 &  20.910 &  19.877 &  18.276 \\ 
cep018 & 0:16:04.10 & -39:16:50.6 &  46.3493 & 2452807.55483 &  21.349 &  20.094 &  18.149 \\ 
cep019 & 0:15:48.48 & -39:15:41.1 &  46.2082 & 2452838.40323 &  20.506 &  19.741 &  18.555 \\ 
cep020 & 0:14:25.63 & -39:09:57.3 &  44.0077 & 2452838.42241 &  21.268 &  20.181 &  18.496 \\ 
cep021 & 0:15:41.65 & -39:15:15.1 &  42.7678 & 2452808.11211 &  20.987 &  19.964 &  18.378 \\ 
cep022 & 0:15:00.42 & -39:12:26.5 &  41.7433 & 2452829.14536 &  21.095 &  20.401 &  19.325 \\ 
cep023 & 0:14:18.63 & -39:10:39.3 &  41.7395 & 2452811.35926 &  21.122 &  20.469 &  19.457 \\ 
cep024 & 0:14:37.93 & -39:10:52.2 &  41.4594 & 2452818.60556 &  20.952 &  20.057 &  18.670 \\ 
cep025 & 0:15:04.49 & -39:12:37.4 &  40.2691 & 2452829.58075 &  20.561 &  19.784 &  18.580 \\ 
cep026 & 0:14:43.45 & -39:11:14.8 &  39.0861 & 2452837.83586 &  21.419 &  20.469 &  18.996 \\ 
cep027 & 0:14:39.76 & -39:10:41.7 &  38.7170 & 2452814.19686 &  21.456 &  20.152 &  18.131 \\ 
cep028 & 0:14:38.17 & -39:11:30.5 &  37.8170 & 2452817.09891 &  21.316 &  20.365 &  18.891 \\ 
cep029 & 0:14:55.46 & -39:13:18.8 &  36.3548 & 2452818.79637 &  20.800 &  20.161 &  19.171 \\ 
cep030 & 0:15:46.52 & -39:16:18.3 &  36.1691 & 2452822.55861 &  21.621 &  20.572 &  18.946 \\ 
cep031 & 0:14:42.58 & -39:11:28.4 &  35.6486 & 2452842.23736 &  20.970 &  20.641 &  20.131 \\ 
cep032 & 0:14:42.52 & -39:11:27.6 &  35.5466 & 2452820.18259 &  19.621 &  18.851 &  17.657 \\ 
cep033 & 0:16:04.21 & -39:16:30.6 &  35.0516 & 2452848.08098 &  21.889 &  20.207 &  17.600 \\ 
\enddata
\end{deluxetable}

\setcounter{table}{1}
\begin{deluxetable}{c c c c c c c c}
\tablecaption{Cepheids in NGC 55 - continued}
%\tablewidth{0pt}
\tablehead{
\colhead{ID} & \colhead{RA} & \colhead{DEC}  & \colhead{P} & \colhead{
${\rm
T}_{0}$} &
\colhead{$<V>$} & \colhead{$<I>$} & \colhead{$<W_{\rm I}>$} \\
 & \colhead{(J2000)} & \colhead{(J2000)}  &
\colhead{ [days]} &  &
\colhead{[mag]} & \colhead{[mag]} & \colhead{[mag]} 
}
\startdata
cep034 & 0:14:44.36 & -39:11:21.5 &  34.5518 & 2452816.49099 &  20.663 &  20.151 &  19.357 \\ 
cep035 & 0:14:53.98 & -39:12:09.2 &  34.1178 & 2452825.84880 &  99.999 &  20.277 &  99.999 \\ 
cep036 & 0:15:11.80 & -39:12:52.4 &  32.8210 & 2452821.22415 &  21.134 &  20.064 &  18.405 \\ 
cep037 & 0:14:25.31 & -39:10:14.9 &  32.4267 & 2452827.27556 &  20.915 &  20.026 &  18.648 \\ 
cep038 & 0:14:59.73 & -39:11:54.7 &  31.5292 & 2452822.25693 &  21.746 &  20.608 &  18.844 \\ 
cep039 & 0:14:19.43 & -39:09:46.7 &  31.1574 & 2452850.00292 &  21.076 &  20.395 &  19.339 \\ 
cep040 & 0:15:02.95 & -39:12:27.5 &  31.0364 & 2452845.00824 &  21.143 &  20.337 &  19.088 \\ 
cep041 & 0:14:58.86 & -39:12:50.0 &  30.2467 & 2452826.35274 &  21.029 &  20.335 &  19.259 \\ 
cep042 & 0:15:33.90 & -39:14:48.5 &  29.5002 & 2452830.30335 &  20.976 &  20.290 &  19.227 \\ 
cep043 & 0:16:02.28 & -39:16:18.2 &  29.4512 & 2452837.34388 &  21.908 &  20.578 &  18.516 \\ 
cep044 & 0:15:54.86 & -39:15:19.3 &  28.0657 & 2452848.53476 &  21.433 &  20.510 &  19.079 \\ 
cep045 & 0:14:54.02 & -39:12:19.3 &  28.0509 & 2452841.89663 &  21.048 &  20.446 &  19.513 \\ 
cep046 & 0:14:38.92 & -39:11:05.1 &  27.7661 & 2452822.84443 &  21.288 &  20.349 &  18.894 \\ 
cep047 & 0:14:42.78 & -39:11:26.6 &  27.2861 & 2452830.27499 &  20.623 &  20.283 &  19.756 \\ 
cep048 & 0:14:50.55 & -39:12:22.7 &  27.1599 & 2452831.66545 &  21.084 &  20.283 &  19.041 \\ 
cep049 & 0:15:40.28 & -39:13:39.8 &  26.6976 & 2452838.50199 &  21.564 &  20.735 &  19.450 \\ 
cep050 & 0:15:34.45 & -39:15:13.7 &  26.3878 & 2452826.21631 &  21.234 &  20.605 &  19.630 \\ 
cep051 & 0:15:37.86 & -39:14:48.6 &  26.3199 & 2452826.63416 &  22.733 &  21.131 &  18.648 \\ 
cep052 & 0:15:30.94 & -39:12:34.0 &  26.0379 & 2452849.86527 &  21.643 &  20.511 &  18.756 \\ 
cep053 & 0:14:46.23 & -39:11:43.5 &  25.5505 & 2452845.03937 &  21.430 &  20.503 &  19.066 \\ 
cep054 & 0:15:02.18 & -39:14:04.5 &  25.4368 & 2452834.77264 &  21.014 &  20.531 &  19.782 \\ 
cep055 & 0:15:16.57 & -39:12:46.6 &  25.1401 & 2452840.13751 &  21.761 &  20.598 &  18.795 \\ 
cep056 & 0:15:59.28 & -39:16:12.6 &  24.9857 & 2452833.12543 &  21.693 &  20.621 &  18.959 \\ 
cep057 & 0:16:04.32 & -39:16:46.6 &  24.5666 & 2452840.29535 &  22.032 &  20.648 &  18.503 \\ 
cep058 & 0:15:30.03 & -39:12:20.3 &  23.9016 & 2452835.76728 &  22.219 &  20.912 &  18.886 \\ 
cep059 & 0:15:53.41 & -39:16:08.4 &  23.7508 & 2452832.09632 &  22.276 &  20.976 &  18.961 \\ 
cep060 & 0:14:44.84 & -39:11:08.7 &  23.7096 & 2452829.85330 &  22.218 &  21.210 &  19.648 \\ 
cep061 & 0:15:52.16 & -39:16:37.4 &  23.5439 & 2452826.41263 &  21.592 &  20.677 &  19.259 \\ 
cep062 & 0:15:33.54 & -39:13:28.7 &  23.5351 & 2452841.77805 &  21.755 &  20.371 &  18.226 \\ 
cep063 & 0:14:42.37 & -39:10:46.0 &  23.3327 & 2452846.32745 &  21.163 &  20.695 &  19.970 \\ 
cep064 & 0:15:55.80 & -39:16:35.7 &  22.9247 & 2452843.72260 &  21.896 &  20.719 &  18.895 \\ 
cep065 & 0:14:46.56 & -39:11:12.9 &  22.8332 & 2452846.49672 &  22.304 &  20.908 &  18.744 \\ 
cep066 & 0:15:32.71 & -39:14:39.4 &  22.6913 & 2452845.71810 &  20.965 &  20.145 &  18.874 \\ 
cep067 & 0:14:34.49 & -39:10:25.3 &  22.4043 & 2452836.82067 &  21.495 &  20.749 &  19.593 \\ 
\enddata
\end{deluxetable}

\setcounter{table}{1}
\begin{deluxetable}{c c c c c c c c}
\tablecaption{Cepheids in NGC 55 - continued}
%\tablewidth{0pt}
\tablehead{
\colhead{ID} & \colhead{RA} & \colhead{DEC}  & \colhead{P} & \colhead{
${\rm T}_{0}$} &
\colhead{$<V>$} & \colhead{$<I>$} & \colhead{$<W_{\rm I}>$}  \\
 & \colhead{(J2000)} & \colhead{(J2000)}  &
\colhead{ [days]} &  &
\colhead{[mag]} & \colhead{[mag]} & \colhead{[mag]} 
}
\startdata
cep068 & 0:14:49.80 & -39:13:23.1 &  22.2934 & 2452844.54875 &  22.006 &  21.177 &  19.892 \\ 
cep069 & 0:14:25.07 & -39:09:10.8 &  22.2824 & 2452846.57642 &  21.753 &  99.999 &  99.999 \\ 
cep070 & 0:15:10.89 & -39:11:59.2 &  22.1505 & 2452847.40035 &  22.468 &  21.136 &  19.071 \\ 
cep071 & 0:15:38.89 & -39:14:56.5 &  22.0950 & 2452833.70839 &  21.252 &  20.785 &  20.061 \\ 
cep072 & 0:15:18.53 & -39:12:29.6 &  21.9627 & 2452837.12408 &  99.999 &  21.118 &  99.999 \\ 
cep073 & 0:14:36.52 & -39:12:19.1 &  21.8533 & 2452832.62534 &  21.937 &  21.036 &  19.639 \\ 
cep074 & 0:15:50.28 & -39:15:46.1 &  21.5309 & 2452831.85886 &  21.263 &  21.303 &  21.365 \\ 
cep075 & 0:14:45.91 & -39:11:27.6 &  21.2743 & 2452842.58589 &  22.084 &  20.868 &  18.983 \\ 
cep076 & 0:16:07.66 & -39:17:28.2 &  21.2415 & 2452846.95235 &  22.539 &  20.933 &  18.444 \\ 
cep077 & 0:14:02.69 & -39:11:09.3 &  20.5319 & 2452831.18683 &  21.189 &  20.648 &  19.809 \\ 
cep078 & 0:15:44.34 & -39:15:12.2 &  20.3943 & 2452835.39236 &  21.673 &  20.826 &  19.513 \\ 
cep079 & 0:14:17.12 & -39:11:41.9 &  20.3942 & 2452845.84048 &  20.952 &  20.166 &  18.948 \\ 
cep080 & 0:14:44.24 & -39:12:45.4 &  20.2633 & 2452845.15643 &  21.504 &  20.859 &  19.859 \\ 
cep081 & 0:15:18.26 & -39:12:14.1 &  19.8080 & 2452842.48408 &  22.407 &  21.007 &  18.837 \\ 
cep082 & 0:15:49.29 & -39:15:48.3 &  19.3776 & 2452838.69568 &  21.852 &  20.947 &  19.544 \\ 
cep083 & 0:14:58.69 & -39:11:54.3 &  19.3580 & 2452847.98353 &  21.926 &  99.999 &  99.999 \\ 
cep084 & 0:14:55.31 & -39:12:50.8 &  19.3436 & 2452836.30750 &  21.131 &  21.072 &  20.981 \\ 
cep085 & 0:14:46.96 & -39:12:37.3 &  19.0477 & 2452834.05731 &  21.426 &  20.924 &  20.146 \\ 
cep086 & 0:15:53.91 & -39:16:25.3 &  18.5752 & 2452836.59181 &  21.745 &  20.478 &  18.514 \\ 
cep087 & 0:14:20.17 & -39:11:24.2 &  18.3966 & 2452847.63746 &  21.764 &  21.053 &  19.951 \\ 
cep088 & 0:15:16.02 & -39:11:53.3 &  18.0795 & 2452834.01755 &  22.852 &  21.235 &  18.729 \\ 
cep089 & 0:14:46.51 & -39:10:42.0 &  17.9405 & 2452835.18692 &  21.352 &  20.453 &  19.060 \\ 
cep090 & 0:14:40.23 & -39:10:03.3 &  17.8206 & 2452847.60431 &  21.811 &  20.837 &  19.327 \\ 
cep091 & 0:14:59.60 & -39:11:48.9 &  17.6201 & 2452847.29662 &  99.999 &  21.131 &  99.999 \\ 
cep092 & 0:14:29.65 & -39:10:16.4 &  17.5669 & 2452837.70808 &  22.679 &  21.178 &  18.851 \\ 
cep093 & 0:15:29.94 & -39:15:06.4 &  17.5499 & 2452833.43163 &  21.819 &  20.561 &  18.611 \\ 
cep094 & 0:15:31.13 & -39:15:40.7 &  17.2671 & 2452847.02598 &  21.770 &  20.835 &  19.386 \\ 
cep095 & 0:14:48.08 & -39:11:25.8 &  17.2338 & 2452842.19519 &  21.809 &  20.881 &  19.443 \\ 
cep096 & 0:15:22.52 & -39:13:31.9 &  17.1511 & 2452833.75601 &  21.982 &  21.323 &  20.302 \\ 
cep097 & 0:14:46.82 & -39:10:51.6 &  17.0303 & 2452839.29756 &  21.417 &  20.918 &  20.145 \\ 
cep098 & 0:14:14.50 & -39:10:42.0 &  16.9912 & 2452844.98131 &  21.759 &  20.945 &  19.683 \\ 
cep099 & 0:14:27.99 & -39:11:46.5 &  16.9451 & 2452849.39100 &  22.094 &  21.401 &  20.327 \\ 
cep100 & 0:14:20.95 & -39:10:52.3 &  16.5556 & 2452836.29047 &  21.822 &  21.572 &  21.184 \\ 
cep101 & 0:14:22.12 & -39:10:50.0 &  16.5403 & 2452848.50372 &  21.261 &  20.851 &  20.215 \\ 
\enddata
\end{deluxetable}

\setcounter{table}{1}
\begin{deluxetable}{c c c c c c c c}
\tablecaption{Cepheids in NGC 55 - continued}
%\tablewidth{0pt}
\tablehead{
\colhead{ID} & \colhead{RA} & \colhead{DEC}  & \colhead{P} & \colhead{
${\rm
T}_{0}$} &
\colhead{$<V>$} & \colhead{$<I>$} & \colhead{$<W_{\rm I}>$} \\
 & \colhead{(J2000)} & \colhead{(J2000)}  &
\colhead{ [days]} &  &
\colhead{[mag]} & \colhead{[mag]} & \colhead{[mag]}
}
\startdata
cep102 & 0:14:49.55 & -39:12:20.4 &  15.9295 & 2452834.61847 &  21.287 &  20.208 &  18.536 \\ 
cep103 & 0:16:00.34 & -39:16:13.5 &  15.8413 & 2452840.15070 &  21.558 &  20.960 &  20.033 \\ 
cep104 & 0:15:48.09 & -39:15:48.6 &  15.4922 & 2452848.08276 &  22.004 &  20.891 &  19.166 \\ 
cep105 & 0:15:39.21 & -39:13:23.9 &  15.3373 & 2452839.24376 &  22.385 &  21.340 &  19.720 \\ 
cep106 & 0:15:50.88 & -39:17:01.9 &  15.0940 & 2452840.27171 &  22.547 &  21.535 &  19.966 \\ 
cep107 & 0:14:26.94 & -39:10:37.8 &  15.0797 & 2452837.95823 &  22.215 &  21.460 &  20.290 \\ 
cep108 & 0:15:15.19 & -39:12:13.9 &  14.9736 & 2452847.98548 &  22.415 &  21.237 &  19.411 \\ 
cep109 & 0:14:48.72 & -39:13:33.0 &  14.9476 & 2452838.39412 &  21.990 &  21.291 &  20.208 \\ 
cep110 & 0:15:24.97 & -39:12:51.2 &  14.6266 & 2452852.25159 &  99.999 &  20.595 &  99.999 \\ 
cep111 & 0:14:33.45 & -39:12:33.2 &  14.3815 & 2452844.42839 &  22.188 &  21.052 &  19.291 \\ 
cep112 & 0:14:34.73 & -39:12:26.8 &  14.1692 & 2452847.42774 &  22.431 &  21.514 &  20.093 \\ 
cep113 & 0:15:58.40 & -39:16:12.1 &  14.1150 & 2452849.96568 &  22.668 &  20.804 &  17.915 \\ 
cep114 & 0:15:49.00 & -39:16:04.5 &  14.0094 & 2452846.51687 &  22.405 &  21.591 &  20.329 \\ 
cep115 & 0:14:30.72 & -39:09:49.6 &  13.9050 & 2452848.72358 &  22.154 &  21.039 &  19.311 \\ 
cep116 & 0:15:16.28 & -39:13:27.1 &  13.5937 & 2452850.69133 &  22.168 &  21.956 &  21.627 \\ 
cep117 & 0:15:25.45 & -39:15:04.2 &  13.4465 & 2452840.17843 &  22.937 &  21.299 &  18.760 \\ 
cep118 & 0:14:28.15 & -39:10:10.0 &  13.2328 & 2452849.85684 &  21.867 &  20.964 &  19.564 \\ 
cep119 & 0:15:46.71 & -39:16:35.1 &  13.0498 & 2452847.75940 &  22.813 &  99.999 &  99.999 \\ 
cep120 & 0:14:27.44 & -39:12:09.3 &  13.0361 & 2452849.11512 &  21.542 &  21.077 &  20.356 \\ 
cep121 & 0:14:08.26 & -39:12:23.1 &  12.4816 & 2452849.92789 &  22.363 &  21.678 &  20.616 \\ 
cep122 & 0:16:04.38 & -39:16:19.6 &  12.4326 & 2452850.09452 &  22.454 &  21.499 &  20.019 \\ 
cep123 & 0:15:41.26 & -39:14:57.7 &  12.3569 & 2452848.82971 &  21.817 &  21.271 &  20.425 \\ 
cep124 & 0:14:57.99 & -39:12:37.2 &  12.0533 & 2452851.65153 &  21.621 &  21.122 &  20.349 \\ 
cep125 & 0:14:42.00 & -39:12:31.5 &  11.9959 & 2452851.18123 &  21.933 &  21.423 &  20.632 \\ 
cep126 & 0:14:16.76 & -39:09:09.9 &  11.9130 & 2452840.80674 &  22.553 &  21.662 &  20.281 \\ 
cep127 & 0:14:34.94 & -39:12:12.3 &  11.4633 & 2452842.70761 &  22.121 &  21.197 &  19.765 \\ 
cep128 & 0:16:11.77 & -39:16:17.3 &  11.0061 & 2452849.74823 &  22.051 &  21.401 &  20.393 \\ 
cep129 & 0:15:57.70 & -39:16:49.4 &  10.8421 & 2452852.89578 &  22.755 &  21.686 &  20.029 \\ 
cep130 & 0:16:13.34 & -39:16:49.0 &  10.7971 & 2452842.40920 &  22.413 &  21.585 &  20.302 \\ 
cep131 & 0:14:19.89 & -39:11:16.4 &  10.6378 & 2452841.66009 &  22.031 &  21.115 &  19.695 \\ 
cep132 & 0:16:06.00 & -39:15:46.3 &  10.2913 & 2452846.29574 &  22.733 &  21.809 &  20.377 \\ 
cep133 & 0:15:14.60 & -39:13:20.3 &  10.2782 & 2452852.15148 &  22.032 &  20.869 &  19.066 \\ 
\enddata
\end{deluxetable}

\setcounter{table}{1}
\begin{deluxetable}{c c c c c c c c}
\tablecaption{Cepheids in NGC 55 - continued}
%\tablewidth{0pt}
\tablehead{
\colhead{ID} & \colhead{RA} & \colhead{DEC}  & \colhead{P} & \colhead{
${\rm
T}_{0}$} &
\colhead{$<V>$} & \colhead{$<I>$} & \colhead{$<W_{\rm I}>$} \\
 & \colhead{(J2000)} & \colhead{(J2000)}  &
\colhead{ [days]} &  &
\colhead{[mag]} & \colhead{[mag]} & \colhead{[mag]}
}
\startdata
cep134 & 0:14:29.79 & -39:12:24.6 &   9.1664 & 2452848.79550 &  22.448 &  21.789 &  20.768 \\ 
cep135 & 0:14:51.96 & -39:11:44.5 &   8.8605 & 2452853.86214 &  19.997 &  19.931 &  19.829 \\ 
cep136 & 0:13:55.94 & -39:10:57.1 &   8.8481 & 2452847.89737 &  22.383 &  22.206 &  21.932 \\ 
cep137 & 0:16:08.89 & -39:15:26.8 &   8.2505 & 2452841.86954 &  22.710 &  21.937 &  20.739 \\ 
cep138 & 0:14:19.33 & -39:10:36.4 &   7.9824 & 2452843.03592 &  22.466 &  21.714 &  20.548 \\ 
cep139 & 0:16:25.33 & -39:16:27.7 &   7.7652 & 2452848.40174 &  22.798 &  22.382 &  21.737 \\ 
cep140 & 0:14:35.99 & -39:12:22.9 &   7.3029 & 2452846.30757 &  22.303 &  21.148 &  19.358 \\ 
cep141 & 0:15:12.74 & -39:14:54.5 &   6.6286 & 2452845.48814 &  99.999 &  22.250 &  99.999 \\ 
cep142 & 0:16:00.27 & -39:14:59.4 &   6.1346 & 2452846.62733 &  22.869 &  22.188 &  21.132 \\ 
cep143 & 0:14:40.81 & -39:11:35.3 &   5.5692 & 2452848.85807 &  22.832 &  21.525 &  19.499 \\ 
\enddata
\end{deluxetable}

\begin{deluxetable}{ccccc}
\tablecaption{Individual V and I Observations}
\tablehead{
\colhead{object}  & \colhead{filter} &
\colhead{HJD}  & \colhead{mag}  & \colhead{$\sigma_{mag}$}\\
}
\startdata
cep001 & V & 2452850.80482 &  19.242 &   0.008\\
cep001 & V & 2452853.83449 &  19.249 &   0.007\\
cep001 & V & 2452858.82639 &  19.283 &   0.008\\
cep001 & V & 2452860.77519 &  19.261 &   0.013\\
cep001 & V & 2452861.92803 &  19.277 &   0.012\\
cep001 & V & 2452868.74762 &  19.320 &   0.012\\
cep001 & V & 2452872.73327 &  19.319 &   0.008\\
cep001 & V & 2452880.74169 &  19.374 &   0.009\\
cep001 & V & 2452886.72905 &  19.415 &   0.013\\
cep001 & V & 2452904.67335 &  19.449 &   0.010\\
cep001 & V & 2452912.67723 &  19.439 &   0.010\\
cep001 & V & 2452917.63772 &  19.448 &   0.015\\
cep001 & V & 2452921.63502 &  19.437 &   0.027\\
cep001 & V & 2452923.65303 &  19.429 &   0.021\\
cep001 & V & 2452931.63660 &  19.420 &   0.010\\
cep001 & V & 2452935.60175 &  19.396 &   0.011\\
cep001 & V & 2452944.69719 &  19.341 &   0.014\\
cep001 & V & 2452951.55163 &  19.259 &   0.021\\
cep001 & V & 2452956.54067 &  19.233 &   0.008\\
cep001 & V & 2452970.52595 &  19.104 &   0.009\\
\enddata
\tablecomments{The complete version of this table is in the electronic
edition of the Journal.  The printed edition contains only
the the first 20 measurements in the V band for the Cepheid variable cep001.}

\end{deluxetable}

\begin{figure}[p]
\vspace*{18cm}
\includegraphics{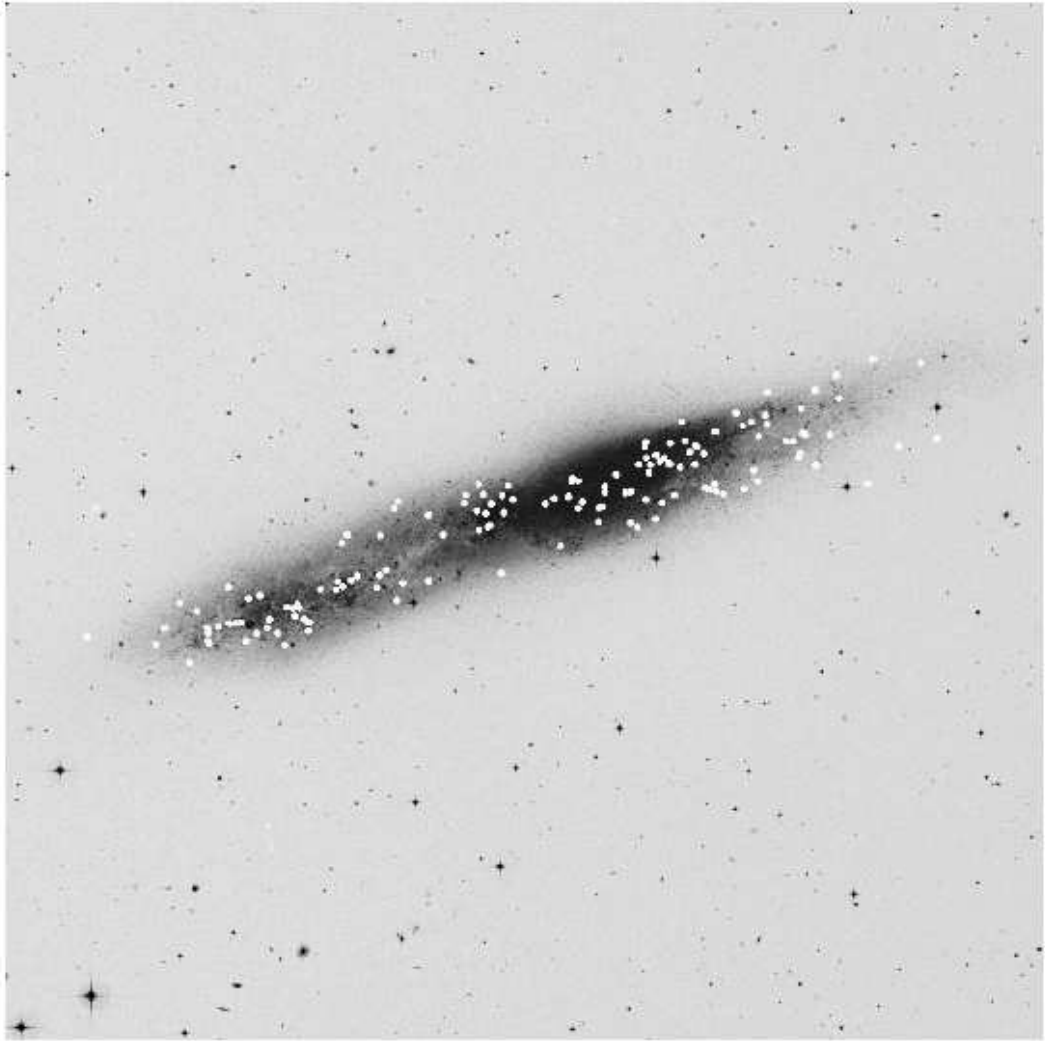}
\caption{The location of the observed field in NGC 55 on the DSS
blue plate. The field of view was about 35 x 35 arcmin. North is up and
East is to the left. The locations of the Cepheids discovered in our survey
are indicated by the white circles.}
\end{figure}

\begin{figure}[htb]
\vspace*{22cm}
\includegraphics{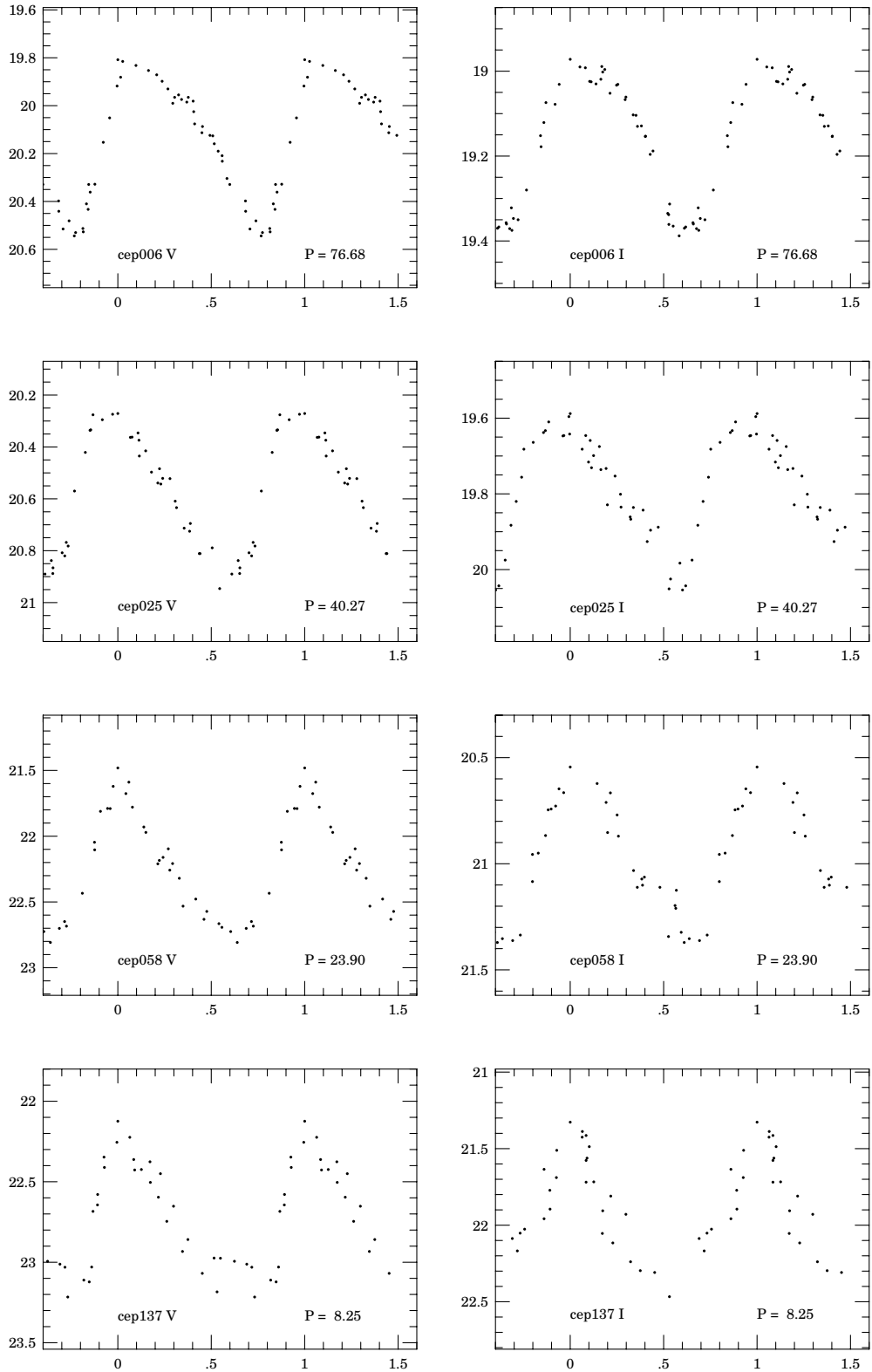}
\caption{Phased V- and I-band light curves for some Cepheids of different
periods in our NGC 55 catalog. These light curves are representative for the
light curves of other Cepheid variables of similar periods.
}
\end{figure}

\begin{figure}[htb]
\vspace*{15cm}
\includegraphics{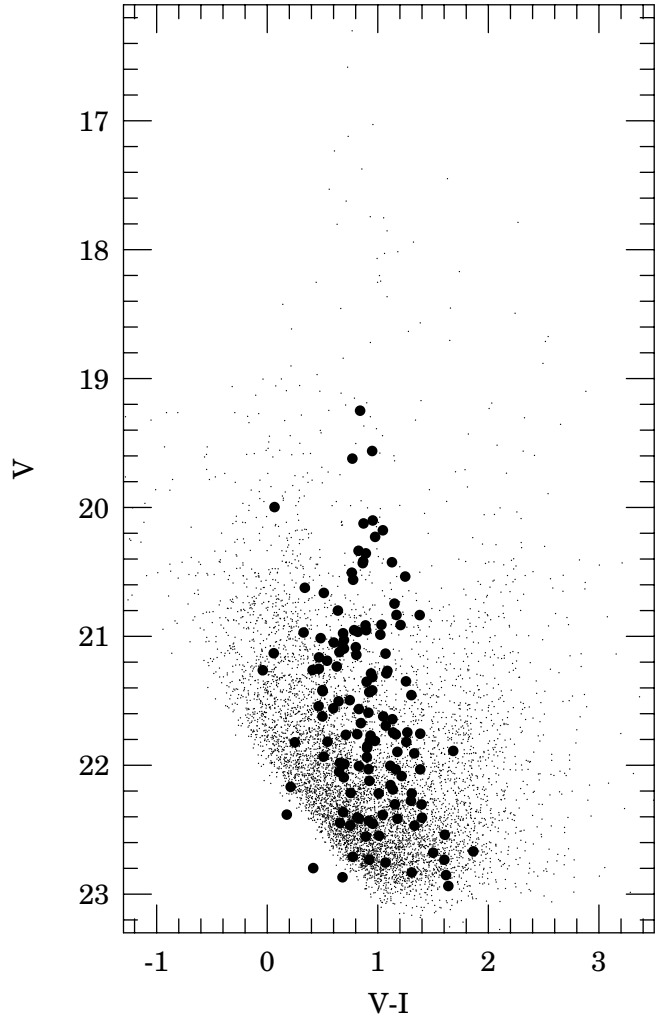}
\caption{The V,V-I color-magnitude diagram for NGC 55. The Cepheids discovered
in our survey are marked with filled circles. 
}
\end{figure}

\begin{figure}[htb]
\vspace*{15cm}
\includegraphics{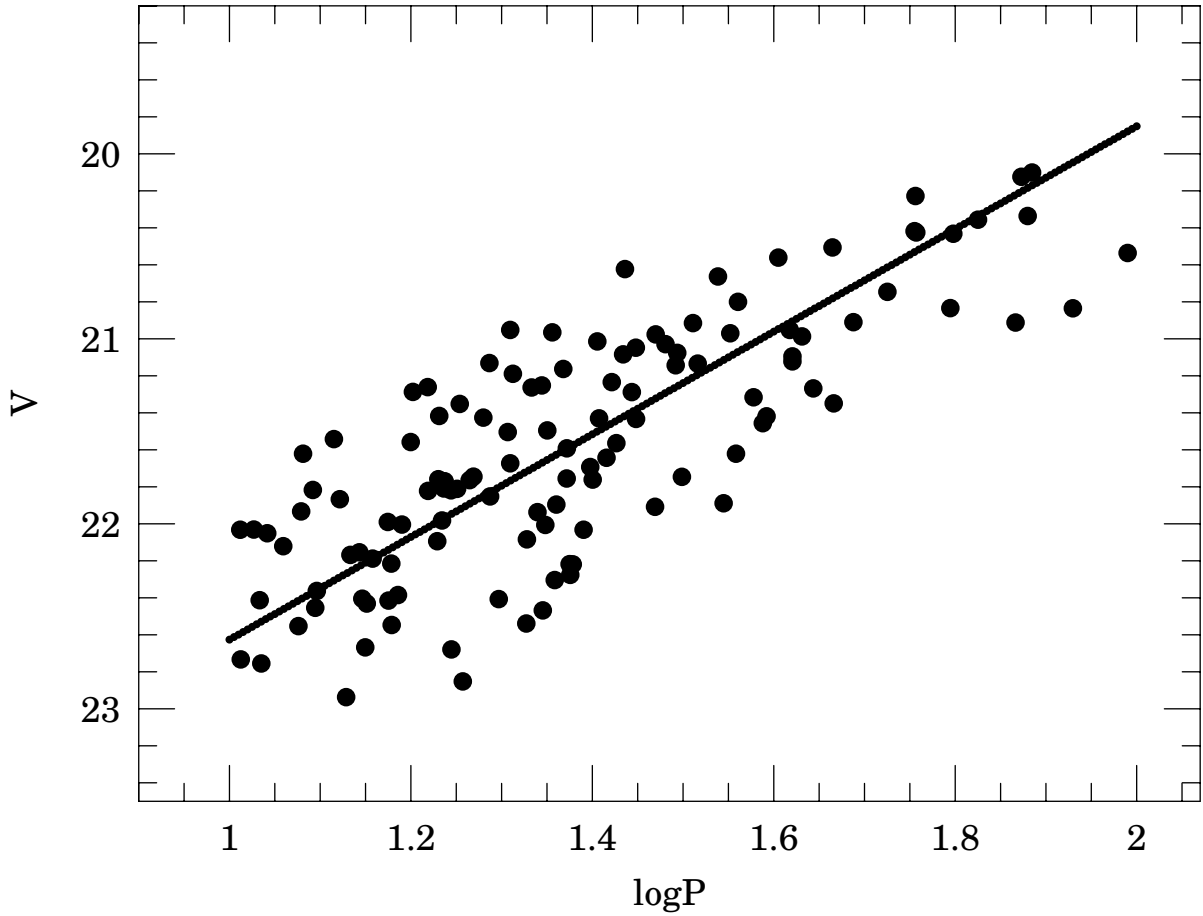}
\caption{The period-luminosity relation for NGC 55 Cepheids in the V 
band. The slope of the relation has been adopted from the LMC Cepheids (see text).
We have used the 130 Cepheids with periods between 10 and 100 days for the
distance determination to NGC 55.
}
\end{figure}

\begin{figure}[htb]
\vspace*{15cm}
\includegraphics{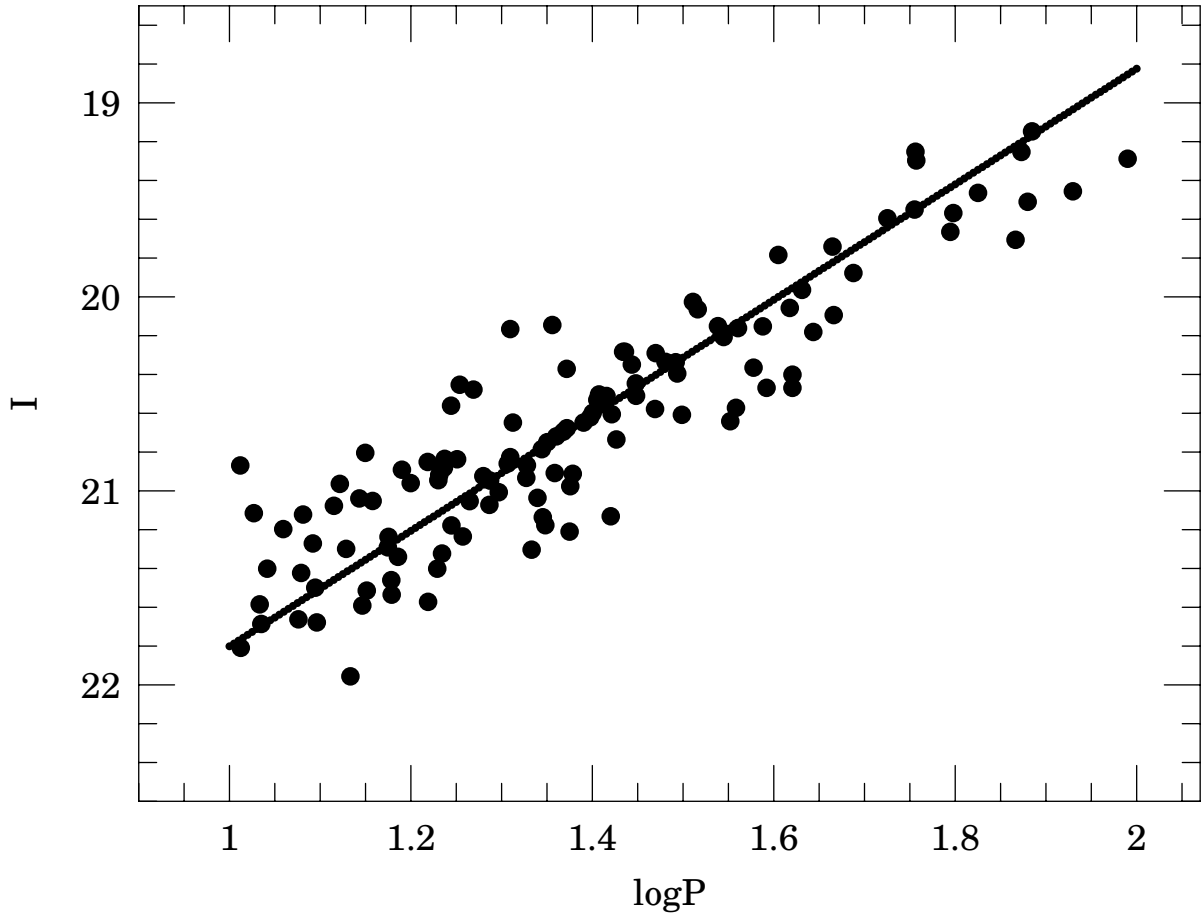}
\caption{Same as Fig. 4, for the I band.
}
\end{figure}

\begin{figure}[htb]
\vspace*{15cm}
\includegraphics{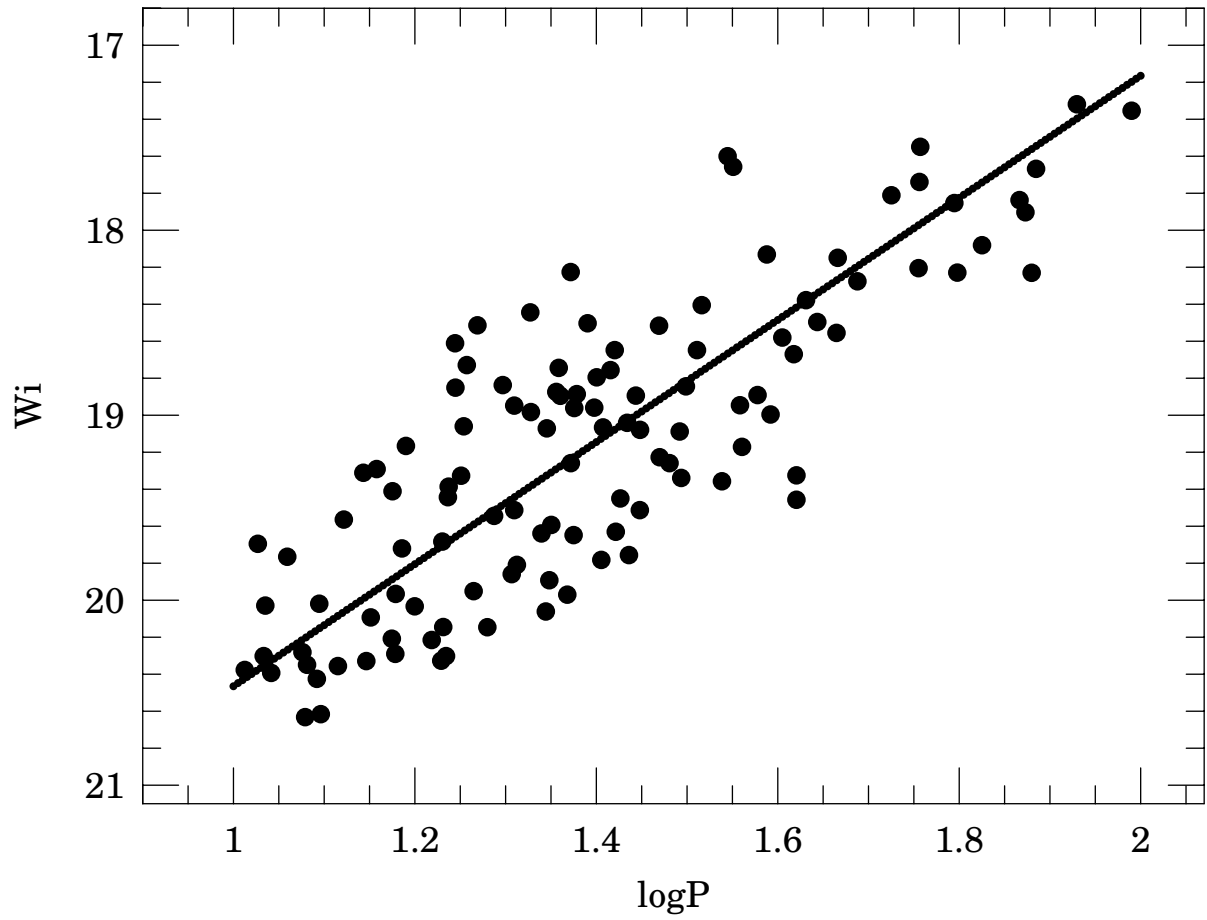}
\caption{Same as Fig. 4, for the reddening-independent (V-I) Wesenheit 
magnitudes. 
}
\end{figure}

\end{document}